\documentclass{Interspeech}

\interspeechcameraready

\title{Identifying Primary Stress Across Related Languages and Dialects with Transformer-based Speech Encoder Models}

\author[affiliation={1,2,3}]{Nikola}{Ljubešić}
\author[affiliation={1}]{Ivan}{Porupski}
\author[affiliation={1}]{Peter}{Rupnik}

\affiliation{Department of Knowledge Technologies}{Jožef Stefan Institute}{Slovenia}
\affiliation{Faculty of Computer and Information Science}{University of Ljubljana}{Slovenia}
\affiliation{Institute of Contemporary History}{Ljubljana}{Slovenia}
\email{nikola.ljubesic@ijs.si, ivan.porupski@ijs.si, peter.rupnik@ijs.si}
\keywords{primary stress detection, pre-trained encoder models, South Slavic languages}

\usepackage{comment}
\usepackage{todonotes}
\begin{document}

\maketitle

\begin{abstract}


  Automating primary stress identification has been an active research field due to the role of stress in encoding meaning and aiding speech comprehension. Previous studies relied mainly on traditional acoustic features and English datasets. In this paper we investigate the approach of
  fine-tuning a pre-trained transformer model with an audio frame classification head. Our experiments use a new Croatian training dataset, with test sets in Croatian, Serbian, the Chakavian dialect, and Slovenian.

  By comparing an SVM classifier using traditional acoustic features with the fine-tuned speech transformer, we demonstrate the transformer’s superiority across the board, achieving near-perfect results for Croatian and Serbian, with a 10-point performance drop for more distant Chakavian and Slovenian. Finally, we show that only a few hundred multi-syllabic training words suffice for strong performance. We release our datasets and model under permissive licenses.



\end{abstract}

\section{Introduction}

Primary stress is a feature of each multi-syllabic word, where one syllable is perceived to stand out from its environment~\cite{tepperman2005automatic}. It has an important and varying function in different languages, including distinguishing word meaning and function, aiding speech comprehension, as well as communicating various sociolinguistic cues~\cite{garde1993naglasak, skaric2007fonetika, subotic2012fonetika}.


Automating the identification of the position of primary stress has attracted significant amount of previous work. Traditionally, the task was performed via supervised learning over acoustic features, while recently transformers pre-trained on speech started to be used, but only for probing experiments~\cite{bentum2024processing} and feature extraction~\cite{mallela2024exploring}. Most of the intended use cases and datasets are related to computer-assisted language learning~\cite{tepperman2005automatic,li2011prominence}, speech synthesis~\cite{verma2006word}, and some applications in research of children's speech~\cite{shahin2014classification, mckechnie2021automated}. The vast majority of work has been performed on English, with very infrequent exceptions such as German~\cite{vakil2015automatic} and Arabic~\cite{shahin2016automatic}.


In this work, we investigate the capacity of pre-trained speech transformer models to predict the position of the primary stress in a multi-syllabic words. We achieve that by fine-tuning the transformer network on that task. Our imminent goal is to apply the resulting technology to describe the variation in spoken language by annotating large spoken corpora, but we also plan more specific downstream use cases, including child language and atypical speech processing, as well as language learning. We break from the tradition of English-centric research by building one training and four test sets in various South Slavic languages and dialects. With this setup, we also investigate the transferability of the proposed technology to similar languages and dialects.

Our training and one of the test languages is Croatian, particularly interesting for the task due to its high dialectal variability, which results in a varying position of the stress even in official communication. Croatian spreads across the Shtokavian, the Kajkavian and the Chakavian dialectal group. Serbian, our test language, mutually intelligible with standard Croatian, has less variability due to its dominant dialect, Shtokavian, which was used in the standardization of both Croatian and Serbian. Slovenian, another of our test languages, is standardized somewhat closely to the Croatian Kajkavian dialect. Finally, our test dialect is the Chakavian dialect, not present in the Croatian standard, nor as close to the Slovenian standard as Kajkavian~\cite{skaric2007fonetika, subotic2012fonetika, greenberg2006short}.

Our primary contributions are the following: (1) we show excellent performance of the pre-trained speech transformers fine-tuned to the task of primary stress identification, (2) we investigate the limitations of the technology when applied to related languages and dialects, (3) we show that supervision of pre-trained models on a few hundred words already yields comparable results to those obtained after fine-tuning the model on ten thousand words, (4) we release a new training dataset and four test sets for related South Slavic languages and dialects, as well as a strong model for Croatian and Serbian\footnote{Data and model are available at \\\url{https://doi.org/10.57967/hf/5658}.}.    

\section{Related work}

The traditional way of performing primary stress identification has been to use prosodic features, such as nucleus duration, intensity and pitch (F0)~\cite{tepperman2005automatic, li2011prominence, shahin2016automatic}, as well as the sonority-based TCSSBC feature contour~\cite{narayanan2005speech, yarra2017automatic}.

These features are mostly exploited in a supervised machine learning setup, with infrequent takes on unsupervised approaches~\cite{ramanathi2019asr}. Most approaches use pre-neural classifiers such as Gaussian Mixture Models~\cite{tepperman2005automatic, verma2006word, li2011prominence}, Hidden Markov Models~\cite{lai2006hierarchical} and Support Vector Machines~\cite{verma2006word, yarra2017automatic}, with some neural exceptions~\cite{shahin2014classification, mckechnie2021automated, shahin2016automatic}. The performance reported from these experiments on various datasets ranges from 72 to 93\% word-level accuracy.

Recently, pre-trained speech transformer models have been applied on the task as well, but only either for probing these networks for primary stress signal~\cite{bentum2024processing}, or for feature extraction for various traditional and neural classifiers~\cite{mallela2024exploring}.


The probing experiments in~\cite{bentum2024processing} report promising results for the availability of the primary stress signal in neural representations, showing that CNN feature extractors already contain a level of relevant signal similar to traditional acoustic features, with higher transformer layers being significantly more predictive of the phenomenon.

The classification experiments in~\cite{mallela2024exploring} compare traditional acoustic features with syllable-averaged neural representations from pre-trained speech transformers, with neural representations being more informative for the task. These experiments also compare traditional and neural classifiers, showing that the latter are more potent on the task. They, however, miss on the opportunity to fine-tune the transformer model to the task directly, which also introduces information loss while averaging the neural representations, available otherwise for each \SI{20}{\milli\second} frame, over the span of each syllable.

\section{Data}

\subsection{Sources}

We construct new training and test datasets by exploiting recently released open datasets in three South Slavic languages and one dialect.

The Croatian \texttt{ParlaStress-HR} training and test datasets are constructed from a sample of the ParlaSpeech-HR open dataset of sentence-aligned parliamentary recordings and transcripts of the Croatian parliament~\cite{ljubevsic2024parlaspeech}. Transcription sentences are sampled to assure the diversity of speakers and gender balance. We split the dataset into a training portion and a test portion, ensuring no speaker overlap while maintaining gender balance.

The Serbian \texttt{ParlaStress-SR} test set is  built from the ParlaSpeech-RS dataset, another member of the
ParlaSpeech collection of speech and text datasets~\cite{ljubevsic2024parlaspeech}, sampling transcript sentences to ensure maximum diversity of speakers, while ensuring gender balance.

The Chakavian \texttt{MićiPrinc-CKM} test dataset is a sample of two chapters from the printed and audio book of the translation of \textit{Le Petit Prince} into the Chakavian dialect. This multimodal book has recently been released as an open dataset with word-level-aligned text and audio~\cite{ljubesic_2024_13936404}.

The Slovenian \texttt{Artur-SL} test set is sampled from the ARTUR dataset~\cite{verdonik2024strategies}, part of the recently updated GOS corpus of spoken Slovenian~\cite{verdonik2024gos}. Three speakers are sampled, one from a public, another from a private setting, and a third one from the parliamentary setting.

\subsection{Data pre-processing}

To enable the manual annotation and subsequent modeling on the level of syllable nuclei, we perform a grapheme-level alignment on all three data sources except \texttt{Artur-SL}, which already had grapheme-level alignment present~\cite{krivzaj2024utilizing}. Phonemic transcription is not needed, except for three simple replacement rules that cover digraphs, due to the usage of phonemic orthography in all the languages and dialects addressed. We align the three datasets by using the forced alignment model from the Kaldi toolkit~\cite{povey2011kaldi} that has previously been released as part of the initial ParlaSpeech dataset construction
efforts~\cite{ljubevsic2022parlaspeech}.

\subsection{Manual data annotation}

Each dataset is annotated by a native speaker using Praat~\cite{Boersma2001} TextGrids, with syllable nuclei of multi-syllabic words pre-selected as candidates for annotation. The annotators are instructed to select one of the syllable nuclei in each multi-syllabic word as the primary stress. In rare cases of deviating transcripts or alignment errors, annotators are instructed to label them with dedicated symbols, and they are not included in the final dataset.

To measure the subjectivity of the task at hand and perform quality control over the obtained manual annotations, we double annotate
the whole \texttt{MićiPrinc-CKM} dialectal test set, given the general agreement among the annotators and authors that primary stress is the hardest to determine in that dataset.
We obtain a high observed word-level agreement of 96.2\% and Krippendorff $\alpha$~\cite{krippendorff2011computing} of 0.92, which proves the quality of our annotations, but also the straightforwardness of the task for humans. Such high levels of inter-annotator agreement on language data are otherwise very rarely observed~\cite{klie2024analyzing}. 

The final size of each of the datasets in terms of the number of syllables, multi-syllabic words and speakers is given in Table~\ref{tab:datasets}. 
The size of the training dataset follows similar English datasets, such as ISLE~\cite{menzel2000isle},
while our test sets are large enough for a reasonable performance estimate of various models, as proven by confidence intervals reported in Section~\ref{sec:results}.

\begin{table}[th]
  \caption{Overview of the size of the train and the test datasets used. Dataset suffixes encode language or dialect (HR~Croatian, SR Serbian, CKM Chakavian, SL Slovenian).}
  \label{tab:datasets}
  \centering
  \begin{tabular}{l r r r }
    \toprule
    \multicolumn{1}{c}{\textbf{Dataset}}   &
    \multicolumn{1}{c}{\textbf{Syllables}} &
    \multicolumn{1}{c}{\textbf{Words}}     &
    \multicolumn{1}{c}{\textbf{Speakers}}
    \\
    \midrule
    \multicolumn{4}{c}{Training dataset}
    \\
    \midrule
    ParlaStress-HR                         & 30561 & 10443 &
    46                                                       \\
    \midrule
    \multicolumn{4}{c}{Test datasets}
    \\
    \midrule
    ParlaStress-HR                         & 3843  & 1291  &
    8                                                        \\
    ParlaStress-SR                         & 1766  & 580   &
    40                                                       \\
    MićiPrinc-CKM                          & 760   & 324   &
    4                                                        \\
    Artur-SL                               & 382   & 136   &
    3                                                        \\
    \bottomrule
  \end{tabular}

\end{table}

\subsection{Data analysis\label{sec:analysis}}

Before moving on to the use of these datasets in machine learning experiments, we performed two short analyses to gain a better understanding of the newly developed datasets.

The first analysis is directed at measuring the variation in the position of the stress in identically spelled words inside Croatian training data, which is one of the main motivations for this work. Already in our training dataset of slightly more than 10 thousand words, if we consider words that occur at least five times, around 6\% of words have a varying position of the primary stress. By inspecting these words manually, we confirmed that the variation is not due to a different part-of-speech category or homography, but rather due to the expected variation in pronunciation of the same words. 

The second analysis aims to measure the similarity of the three cross-lingual test sets to the Croatian training dataset in terms of the position of the primary stress in identically spelled words. The \texttt{ParlaStress-SR} dataset has an expected and significant lexical overlap of 305 words (53\%), only 6 (2\%) of them having an stress position not observed in the training data. The \texttt{MićiPrinc-CKM} dataset has 63 identical words, 9 (14\%) have a stress position not covered in the training data, showing a more significant deviation of the stress position than the Serbian dataset. Finally, the \texttt{Artur-SL} dataset has only 20 words covered by the training data, but 13 (65\%) of these have a different position of the primary stress, accentuating the large differences in the stress positions between Croatian and Slovenian.


\section{Methods}

\subsection{Pre-trained transformer model\label{sec:transformer}}

Our solution for the problem at hand is the fine-tuning of the w2v-bert-2.0 model\footnote{\url{https://huggingface.co/facebook/w2v-bert-2.0}} with an audio frame classification head on top of the transformer model~\cite{wolf-etal-2020-transformers}, allowing for every \SI{20}{\milli\second} frame to be classified into a specific category. 
The raw transformer model is a 580-million-parameters conformer model which was pre-trained on \SI{4.5}{\mega\hour} of unlabeled audio data covering more than 143 languages, and has shown state-of-the-art results in speech translation and transcription tasks, especially on less-resourced languages~\cite{wav2vec2bertmodelpaper}.

We transform each multi-syllabic word into a sequence of \SI{20}{\milli\second} frames, every frame being labeled as 0, except during the manifestation of the nucleus of the stressed syllable, where the label is 1. With this, we set our problem as a binary classification task on each audio frame.


Optimal hyperparameters are identified based on the hyperparameter search on our training data. We have investigated learning rates of \{\num{8e-6}, \num{1e-5}, \num{3e-5}, \num{5e-5}\}, number of epochs ranging from 1 to 20, and the impact of 1 or 4 gradient accumulation steps. These preliminary experiments showed that the learning rate of \num{1e-5}, no gradient accumulation, and 20 epochs, our batch size being 32, deliver highly stable results on various splits of our training set. We fine-tune our model on an NVIDIA A100 with \SI{40}{\giga \byte} of memory. Fine-tuning for one epoch takes \SI{4.5}{\minute}.

\subsection{Support Vector Machine model\label{sec:svm}}

To compare our approach with traditional methods, we train an SVM model with an RBF kernel and $C=10$, using prosodic features. We consider each syllable nucleus as an instance, for which we calculate the prominence of intensity, pitch, and sonority by dividing the nucleus area under the curve (AUC), mean, and peak values by the corresponding word-level mean. This results in a total of nine input features, plus syllable nucleus duration. Binary classification is performed on each syllable nucleus.

For this classifier we also performed a number of hyperparameter search and feature selection experiments across the training dataset, the hyperparameters ranging between an RBF and linear kernel, and $C$ selected from $\{0.1,1,10,100\}$.



\subsection{Evaluation}

We evaluate each model on word-level accuracy. The predictions of each model are post-processed to ensure that only one syllable nucleus per word is selected as the most likely position of the primary stress.

In case of the transformer classifier, the syllable nucleus closest to the longest range of a span of positive predictions is selected as the final prediction. In very infrequent cases, multiple spans within a word are predicted to be primary stress positions.

For support vector machines, the syllable with the highest positive-class probability on the word level is selected as the final prediction.

\section{Results\label{sec:results}}

\begin{figure*}[t]
  \centering
  \includegraphics[width=\linewidth]{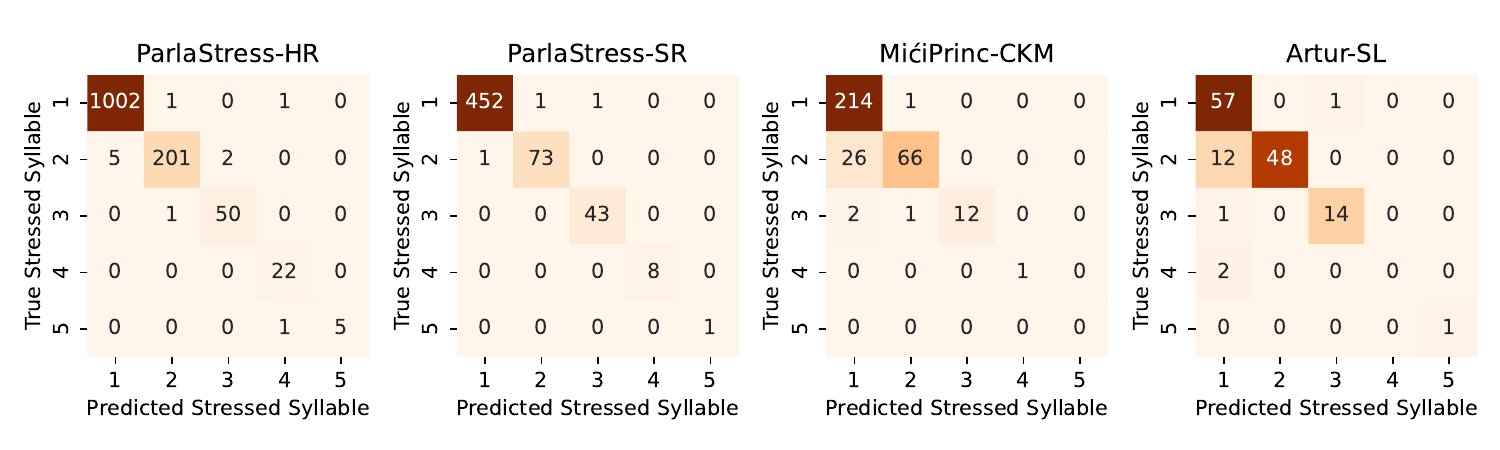}
  \caption{Confusion matrices of stress positions on the four test sets.}
  \label{fig:cm}
\end{figure*}

\begin{figure*}[t]
  \centering
  \includegraphics[width=\linewidth]{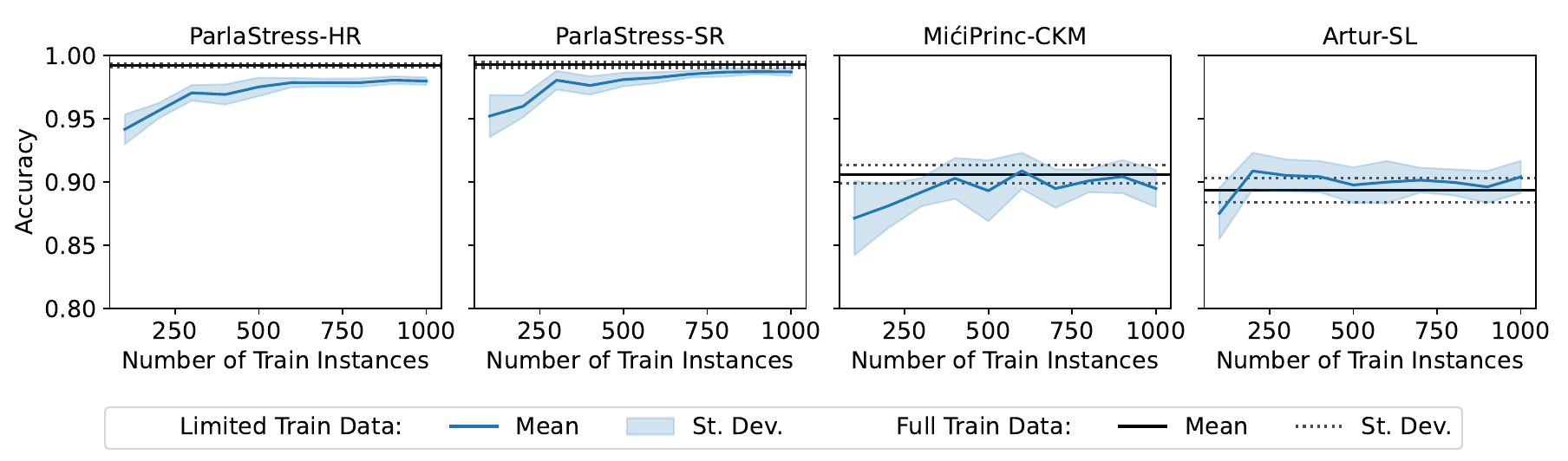}
  \caption{Learning curves as the number of training instances increases, compared to the performance of the final transformer model trained on all available instances.}
  \label{fig:lc}
\end{figure*}

\subsection{Traditional vs. deep features}

In the first experiment, we compare the performance of the SVM classifier, trained on traditional acoustic features, with that of the transformer model. Each classifier was trained on the whole training dataset. The results of each classifier are given in Table~\ref{tab:results}.

The results show significant dominance of the transformer models over the SVM models, with word accuracy differences between 11 and 25 percentage points. However, a significant robustness of the traditional features can be observed as well, with a significantly smaller difference between the results on the various test sets, regardless of the distance to the training data, as discussed in Section~\ref{sec:analysis}.

Both the Croatian and Serbian test sets show to be relatively simple for both methods, but with a drastic difference in performance of less than one percent of error for the transformer and 20\% or more of error for the SVM.

On the Chakavian and Slovenian test sets, the transformer model achieves ten to twelve points lower performance than on the two other datasets. This is in line with the findings in Section~\ref{sec:analysis}, which show a high level of similarity between Croatian and Serbian, and a decreasing similarity of Chakavian, followed by Slovenian. However, regardless of this drop in performance, and the robustness of SVMs across the test sets, transformers still outperform SVMs with more than 10 accuracy points difference.

\begin{table}[th]
  \caption{Comparison of the word-level accuracies of the SVM and the transformer model with 95\% confidence intervals.}
  \label{tab:results}
  \centering
  \begin{tabular}{l r@{\hskip 1em} r r@{\hskip 1em} r}
    \toprule
    \multicolumn{1}{c}{\textbf{Dataset}} &
    \multicolumn{2}{c}{\textbf{SVM}}     &
    \multicolumn{2}{c}{\textbf{w2v-bert-2.0}}
    \\
    \midrule
                                         & \multicolumn{1}{c}{\textbf{acc}}    &
    \multicolumn{1}{c}{\textbf{95\%CI}}  &
    \multicolumn{1}{c}{\textbf{acc}}     & \multicolumn{1}{c}{\textbf{95\%CI}}                                    \\
    \midrule
    ParlaStress-HR                       & 74.0                                & [71.6,76.3] & 99.1 & [98.6,99.6] \\
    ParlaStress-SR                       & 80.2                                & [77.1,83.3] & 99.3 & [98.6,99.8] \\
    MićiPrinc-CKM                        & 78.7                                & [73.8,83.0] & 88.9 & [85.2,92.3] \\
    Artur-SLO                            & 72.1                                & [64.0,79.4] & 89.0 & [83.1,94.1] \\
    \bottomrule
  \end{tabular}

\end{table}


\subsection{Stress position}

In this set of experiments we investigate whether the decreasing performance of transformer models on the Chakavian and Slovenian test sets are due to the model's bias towards a specific stress position introduced by fine-tuning on the Croatian dataset. We calculate confusion matrices between the true and the predicted syllable position from our transformer evaluation results on each of the four test sets. The results are depicted in Figure~\ref{fig:cm}.

In the Croatian and Serbian test set there is an obvious preference for the first syllable of 78\% cases, while that preference drops to 66\% for Chakavian and 43\% on Slovenian. The wrong predictions on both Chakavian and Slovenian are mostly due to the first-syllable stress being preferred.

Additionally, we performed a short manual qualitative analysis of the erroneously classified words in Chakavian and Slovenian. In the Chakavian dataset the most frequent reason for misclassification was a less clear position of the primary stress, while in the Slovenian dataset the most frequent reason was the classifier's preference for an earlier stress position regardless of the stress being clearly pronounced later in the word.






\subsection{Training data size}

In this final set of experiments we investigate the capacity of the transformer model to perform well on smaller amounts of fine-tuning data. During our previous experiments we have observed very good performance already after a single epoch of fine-tuning, which signals that the more than 10 thousand multi-syllabic words we have at our disposal might not be necessary to obtain our final results. The following insights are especially relevant as they illustrate how one could bring the performance on Chakavian and Slovenian up to the levels of Croatian and Serbian via additional manual data annotation.

We perform experiments on varying the amount of training instances from 100 to 1000, with step of 100. The number of training steps is kept constant at~1200 to control for the amount of updates the transformer has received. On each amount of training data, 10~models are trained on a random selection of the training data. We compare the learning curves with the performance of 10~models trained on all data. The variability of model performance is quantified via standard deviation. The results, given in Figure~\ref{fig:lc}, show that even with a few hundred words available for fine-tuning, performance becomes useful and improves significantly up to a training dataset size of only 500 words.
At that point the final performance on the Chakavian and Slovenian test sets is obtained already, while for the Croatian and Serbian test set, more similar to the training data, there is slow growth continuing even after the 1000 instances depicted here.

\section{Conclusion}

This paper has investigated the performance of pre-trained transformer speech encoders on the task of primary stress identification, comparing them to SVM classifiers trained on traditional acoustic features. The experiments were performed on a newly constructed training and four test datasets in various South-Slavic languages and dialects. Although SVM classifiers show to be more robust to language and dialect change, 
their performance in each setup is drastically lower to that of transformers. On Croatian and Serbian, transformer models achieve close-to-perfect results, with a 10-percent accuracy drop on more distant Chakavian and Slovenian.

Insights in the true and predicted position of the stress in transformer models show that the main reason for the drop in performance on Chakavian and Slovenian is the very strong preference of the first syllable in the Croatian training data. Experiments on the impact of training data size show that 500 annotated words for fine-tuning already generate peak performance in Chakavian and Slovenian, while for Croatian and Serbian, having multiple thousands of fine-tuning instances does improve the results further.

Future work will include developing techniques for model robustness to language and dialect change, as well as more in-depth analyses such as gender and word memorization effects.


\section{Acknowledgements}

This work was supported in part by the Projects ``Spoken Language Resources and Speech Technologies for the Slovenian Language'' (Grant J7-4642), ``Large Language Models for Digital Humanities'' (Grant GC-0002), the Research Programme ``Language Resources and Technologies for Slovene'' (Grant P6-0411), and the Research Infrastructure DARIAH-SI (I0-E007), all funded by the ARIS Slovenian Research and Innovation Agency.

We are especially grateful to our two data annotators, Mirna Potočnjak and Nejc Robida.

\bibliographystyle{IEEEtran}
\bibliography{mybib}

\end{document}